\documentclass[%
reprint,
amsmath,amssymb,aps,prb 
nolongbibliography,
]{revtex4-2}

\usepackage{epsfig}% Include figure files
\usepackage{graphicx}% Include figure files
\usepackage{dcolumn}% Align table columns on decimal point
\usepackage{bm}% bold math
\usepackage[export]{adjustbox}
\usepackage{hyperref}
\usepackage{ulem}
\hypersetup{colorlinks=true, linkcolor=blue, citecolor=blue, filecolor=blue,	urlcolor=blue}

\usepackage{xcolor, soul}
\sethlcolor{red}

\begin{document}

\preprint{preprint}

\title{Influence of anti-ferromagnetic ordering and electron correlation on the electronic structure of MnTiO$_3$} 

\author{Asif Ali}
%\email{asifa@iiserb.ac.in}
\affiliation{Department of Physics, Indian Institute of Science Education and Research Bhopal - Bhopal Bypass Road, Bhauri, Bhopal 462066, India}%

\author{R. K. Maurya}
\affiliation{Department of Physics, Indian Institute of Science Education and Research Bhopal - Bhopal Bypass Road, Bhauri, Bhopal 462066, India}%

\author{Sakshi Bansal}
\affiliation{Department of Physics, Indian Institute of Science Education and Research Bhopal - Bhopal Bypass Road, Bhauri, Bhopal 462066, India}%

\author{B. H. Reddy}
\altaffiliation[Presently at ]{Department of Physics, Government College (A), Rajahmundry 533105, India}
%\email{harinath@gcrjy.ac.in}
\affiliation{Department of Physics, Indian Institute of Science Education and Research Bhopal - Bhopal Bypass Road, Bhauri, Bhopal 462066, India}%

\author{Ravi Shankar Singh}
\email{rssingh@iiserb.ac.in}
\affiliation{Department of Physics, Indian Institute of Science Education and Research Bhopal - Bhopal Bypass Road, Bhauri, Bhopal 462066, India}%

\date{\today}
\begin{abstract}
    
  		Electron correlation and long-range magnetic ordering have a significant impact on the electronic structure and physical properties of solids. Here, we investigate the electronic structure of ilmenite MnTiO$_{3}$ using room temperature photoemission spectroscopy and theoretical approaches within density functional theory (DFT), DFT+$U$ and DFT+dynamical mean field theory (DMFT). Mn 2$p$ (Ti 2$p$) core level photoemission spectra, confirming Mn$^{2+}$ (Ti$^{4+}$) oxidation state, exhibit multiple satellites which are very similar to that of MnO (TiO$_{2}$), suggesting similar strength of various interactions in this system. Valence band spectra collected at different photon energies suggest dominant Mn 3$d$ character in the highest occupied band with a wide insulating gap. DFT(+$U$) correctly predicts the experimentally observed anti-ferromagnetic (AFM) insulating ground state for MnTiO$_3$ where the requirement of a large $U$ to reproduce the experimental values of magnetic moment and band gap signifies the importance of electron correlation. Magnetically disordered paramagnetic (PM) phase could be well captured within DFT+DMFT, which provides an excellent agreement for the experimental band gap, paramagnetic moment, valence band spectra as well as dominant Mn 3$d$ character in the highest occupied band. The calculated spectral function remains largely unaffected and exhibits sharper features in the magnetically ordered AFM phase. We show that the electronic structure of MnTiO$_{3}$ in both the PM and AFM phases can be accurately described within DFT+DMFT.
\end{abstract}
\maketitle
	
%\section*{INTRODUCTION}    
    Ilmenites $M$TiO$_{3}$ ($M$ = Mn, Fe, Co, Ni) have attracted significant attention owing to their unique crystal structure and fascinating dielectric, magnetic and optical properties \cite{BarthPosnjak+1934+265+270, CTO_dielectric, Ishikawa1958_MagneticProMTiO3JPSJ, SatoPRL2020_Magnetochiral, SWChen_XAS_APL2014, TMPan_JAP2001}. The complex interplay between charge, spin, lattice and orbital degrees of freedom leads to technologically important properties with potential applications in various fields such as multiferroics, spintronics, sensors, photocatalysts, \textit{etc} \cite{RIBEIRO2018463, RevModPhys.AFMspintronics, HE20085, APARNA2019319, Photocatalyst, TRUONG201285, HaoWang, ie071619d, FUJII2004}. Recent studies revealed magnetoelectric, magnetoelastic, and magnetostructural couplings, as well as the existence of Dirac magnons in this class of materials \cite{ PhysRevB.106.174425,Sato2022_PhysRevB.105.094417,PhysRevB.104.014429, PhysRevB.101.195122, PhysRevB.93.104404, PhysRevB.83.104416, MnTiO3_filmME}, providing opportunities for manipulating and controlling the properties by applying external fields.
\begin{figure*}[t]
	\centering
	\includegraphics[width=0.98\textwidth]{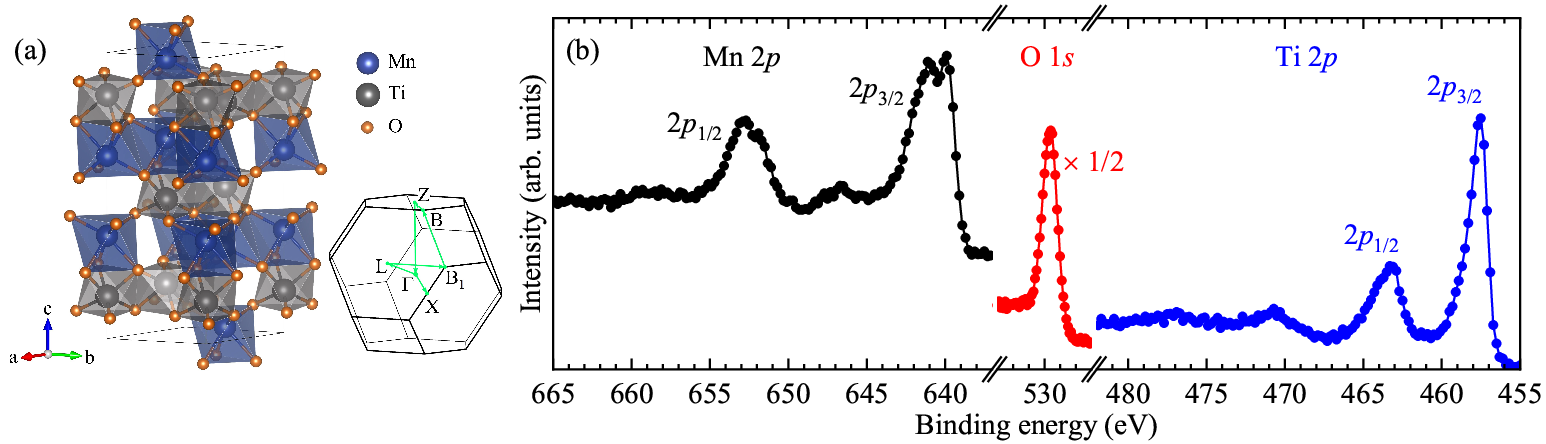}
	\caption{\label{fig:fig1}  (a) Crystal structure of MnTiO$_{3}$ and corresponding Brillouin zone with green lines showing $k$-path along specific high symmetry points. (b) Mn 2\textit{p}, O 1\textit{s} (rescaled by 1/2) and Ti 2\textit{p} core level photoemission spectra collected using Al~$K_{\alpha}$ radiation.}
\end{figure*}
    $M$TiO$_{3}$ belongs to the rhombohedral crystal system comprising alternate layers of buckled honeycomb network along the $c$-axis\cite{BarthPosnjak+1934+265+270}, as illustrated in Fig. \ref{fig:fig1}(a). The magnetic ground state and its complex interplay with various degrees of freedom in $M$TiO$_{3}$ strongly depends on $M$ ion \cite{Ishikawa1958_MagneticProMTiO3JPSJ,  HKato_1986, YOSHIZAWA1992_NiMnTiO3, Yoshizawa_PRL1987_FeMnTiO3, Shirane1959_neutron, Stickler}. 
    In particular, MnTiO$_{3}$, with Ti$^{4+}$ having 3$d^0$ and Mn$^{2+}$ having 3$d^5$ electronic configurations, exhibits collinear anti-ferromagnetic (AFM) ordering below $T_N$ $\approx$ 64 K with the $c$-axis as easy axis \cite{Ishikawa1958_MagneticProMTiO3JPSJ, Shirane1959_neutron}. The $T_N$ is found to be robust down to ultra-thin layer limit \cite{Miura_Tn_MTO_film}. The measured paramagnetic effective moment is close to the expected spin only value of 5.92 $\mu_{B}$/Mn corresponding to $S$ = 5/2 \cite{Stickler, YAMAUCHI1983_spinflop}. Neutron diffraction at 4.2 K reveals G-type AFM ordering with ordered moment of 4.55 $\mu_{B}$/Mn \cite{Shirane1959_neutron}. A broad maxima around 100 K in the bulk magnetic susceptibility indicates in-plane short-range ordering much before $T_N$ along with the observation of spin-lattice coupling in a similar temperature range \cite{Stickler, AKIMITSU197087, Maurya2015}. With observation of strong linear magnetoelectric effect and electric polarization flop along with spin-flop transition, MnTiO$_{3}$ has promising applications such as in multiferroic and storage devices \cite{Scott2007, 1190077, Andreas_Moser_2002}. Substitution of Mn, Ru and Nb at Ti site has been shown to alter the magnetic and electronic properties such as reduction in band gap, increased three dimensional magnetic character and reduced spin-flop magnetic field \cite{PPal_MnDoped, PAR_NbDoped, RKM_RuDoped, RKM_RuDoped2}.

     The band gap in MnTiO$_{3}$ is found to be about 3.4 eV using optical \cite{Enhessari2012} and 2.4 eV using $x$-ray spectroscopic measurements \cite{Agui2011}. Density functional theory (DFT) and DFT+$U$ calculations have been successful in describing the AFM insulating ground state of MnTiO$_{3}$ where the importance of intra-site Coulomb ($U$) and inter-site Hund's ($J_{H}$) interactions have been realized \cite{ SWChen_XAS_APL2014, Maurya_2017, Deng2012, KANG2018251}. However, the electronic structure of the paramagnetic (PM) phase and its evolution to the AFM phase remains largely unexplored, despite several experimental studies in the PM phase \cite{Agui2011,Enhessari2012,SWChen_XAS_APL2014,Agui2015,Maurya_2017}. The electronic structure of magnetic materials significantly changes with magnetic ordering, impacting their physical properties \cite{AKundu_PRB2017_MnOfilm, Kuo2017, Koster_RMP2012_SROreview}. 
	 A few methodologies, such as DFT and molecular dynamics with disordered local moment model \cite{Hubbard_DLM_Iron, JHUbbard_FeDLM2, BLGyorffy_1985, AJPindor_1983, Steneteg_PRB2012_DLMMD, Mozafari_PRB2018_DLMMD_NiO} and dynamical mean field theory (DMFT) \cite{Georges_DMFTRevModPhys.68.13,GKotliar_RevModPhys.78.865}, have been used for proper treatment of thermally fluctuating large local moments in the PM phase \cite{ABRIKOSOV201685}. The combined DFT+DMFT method has emerged as a powerful method to understand the PM as well as magnetic phases of strongly correlated systems \cite{PhysRevB.100.245109, MinjaeKim_PRB2015_DMFTSrRuO3, PhysRevLett.87.067205, Haule_PRL2019_MnPS3, Jonsson_PRB022_MAXphaseDMFT, Tianye_PRB2022_CrI3}.
	
	In this paper, we report a systematic study of the electronic structure of ilmenite MnTiO$_{3}$ using photoemission spectroscopy and theoretical calculations within DFT, DFT+$U$ and DFT+DMFT. Core level photoemission spectra exhibit multiple structures (satellites) due to photoemission final state effects. Valence band spectra collected at different photon energies suggest dominant Mn 3$d$ character in the highest occupied band. Valence band photoemission spectra compared with theoretical results suggests that the DFT+DMFT accurately describes the electronic structure for MnTiO$_{3}$ in the PM phase and it remains largely unaffected by the long-range AFM ordering. 
 
%\section*{Methodology}
   High-quality polycrystalline sample of MnTiO$_{3}$ was prepared by solid state reaction route using stoichiometric amounts of high purity MnO (99.99\%) and TiO$_2$ (99.99\%). A well ground mixture was pelletized and sintered at 1200~$^\circ$C for 48 hours with an intermediate grinding. Room temperature $x$-ray diffraction and its Reitveld refinement reveal hexagonal lattice parameters $a$ = $b$ = 5.138(5) \AA ~and $c$ = 14.284(4) \AA, in good agreement with earlier reports\cite{YAMAUCHI1983_spinflop, Kidoh1984,PhysRevB.83.104416}. The crystal structure and Brillouin zone are shown in Fig. \ref{fig:fig1}(a). Room temperature photoemission measurements were performed on \textit{in-situ} fractured samples using monochromatic Al~$K_\alpha$ ($h\nu$ = 1486.6~eV) and He \textsc{ii} ($h\nu$ = 40.8~eV) radiation sources (energy) with total energy resolution of about  300~meV and 15~meV, respectively. DFT and DFT+$U$ calculations were performed using  \textsc{wien2k}\cite{WIEN2k, Wien2k2019} code. eDMFT code\cite{HauleDMFTPRB2010} with continuous-time quantum Monte Carlo (CTQMC) impurity solver was used for DFT+DMFT calculation. G-type AFM ordering was considered for the AFM phase \cite{Shirane1959_neutron}. More calculation details are in the Supplementary Material (SM).

%\section*{RESULTS AND DISCUSSION}
	Figure \ref{fig:fig1}(b) shows the spectral region of Mn 2$p$, O 1\textit{s} and Ti 2\textit{p} core levels where intensities of all three spectra (along with inelastic background) are matched with the survey scan. The O 1$s$ spectra exhibit a sharp peak at about 529.6 eV binding energy corresponding to lattice oxygen without any additional higher binding energy feature, suggesting clean sample surface.
	Ti 2\textit{p} spectra exhibit two main peaks at about 457.5 eV (2\textit{p}$_{3/2}$) and 463.1 eV (2\textit{p}$_{1/2}$) binding energy with spin-orbit splitting of about 5.6 eV and broad satellite features at about 13.4 eV higher binding energy, corresponding to each main peak. Energy separation between satellite and main peak is predominantly determined by Ti 3$d$ - O 2$p$ hybridization as it affects the screening of the core-hole generated in the photoemission process \cite{PhysRevB.100.075146}. Similar energy separation in  MnTiO$_{3}$ and  rutile TiO$_{2}$ ($\sim$ 13.4 eV) \cite{AHarikiPRB2022_Ti1s2psatellite} and smaller compared to perovskite SrTiO$_{3}$ ($\sim$ 13.9 eV) \cite{AHarikiPRB2022_Ti1s2psatellite}, suggests a similar degree of Ti 3$d$ - O 2$p$ hybridization in MnTiO$_{3}$ and TiO$_{2}$ due to similarity in octahedra connectivity. Interestingly, Ti 2$p$ main peaks exhibit asymmetry towards higher binding energy, as opposed to symmetric peaks in TiO$_2$ and MgTiO$_{3}$ \cite{SAChamber_PRB2017_Ti2O3, MiyoshinoPRB2023_Ti2pMgTiO3}. This asymmetry may arise due to trigonal distortion of the TiO$_{6}$ octahedra and/or due to a direct Mn-Ti interaction possibly arising from face-sharing of MnO$_6$-TiO$_6$ octahedra \cite{MiyoshinoPRB2023_Ti2pMgTiO3,SAChamber_PRB2017_Ti2O3, CFChange_PRX2018_Ti2pdimerTi2O3}. Further investigations are required to uncover the origin of this asymmetry.
	
	In Mn 2\textit{p} spectral region, Mn 2$p_{3/2}$ and Mn 2$p_{1/2}$ appear at about 640 eV and 653 eV binding energy, respectively, with spin-orbit splitting of about 13 eV. Each of the spin-orbit split feature exhibits multiple peaks, as clearly seen for Mn 2$p_{3/2}$, appearing at about 639.9 eV and 641.1 eV binding energy along with shoulder structure leading to asymmetric lineshape towards higher binding energy, while Mn 2$p_{1/2}$ shows a single asymmetric broad feature. Distinct broad satellites are also observed at $\sim$ 7 eV higher binding energy from the main features. The core level satellite along with multiplet features are very similar with respect to the relative intensity and energy position to that obtained in the case of MnO \cite{PhysRevB.73.155110, PhysRevA.63.050702, OkadacorelevelJPSJ_1992}, suggesting similar strength of interaction parameters such as crystal field and Coulomb interaction.  
	
\begin{figure}[t]
		\centering
		\includegraphics[width=0.48\textwidth]{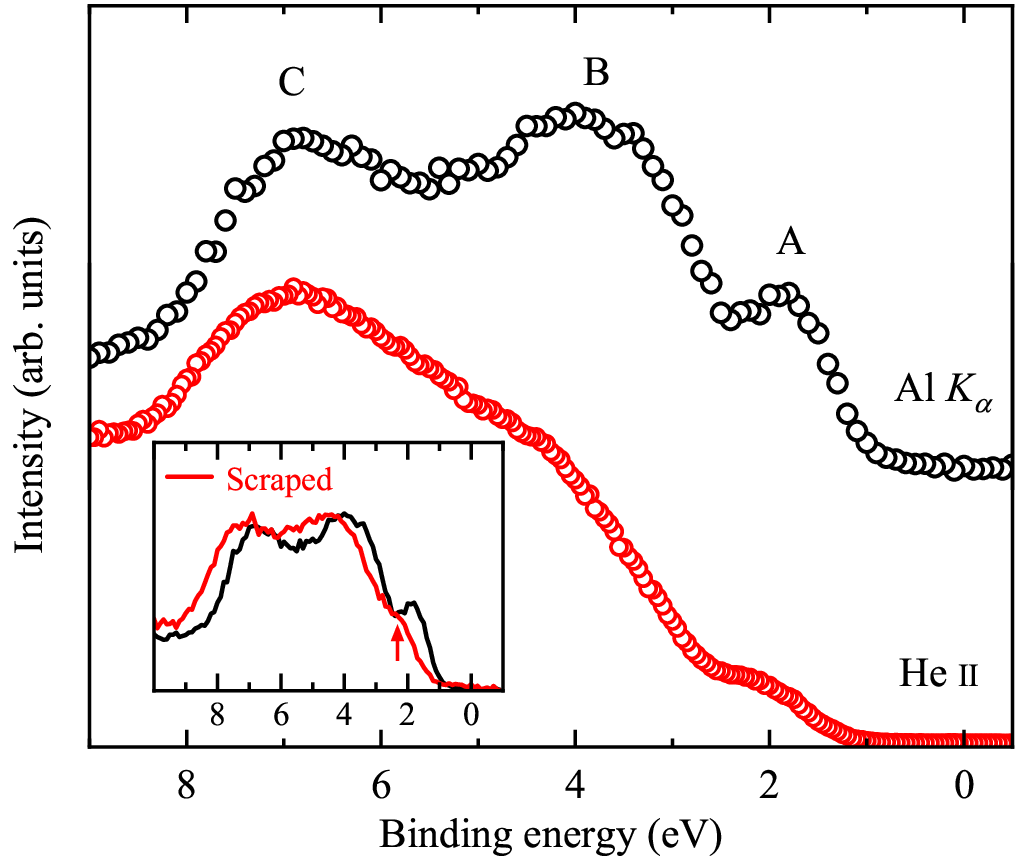}
		\caption{\label{fig:XPS}  Valence band photoemission spectra of MnTiO$_{3}$ collected using Al~$K_\alpha$ and He \textsc{ii} radiations. Inset shows comparison of Al~$K_{\alpha}$ spectra for fractured (black line) and scraped (red line) sample surface. Arrow marks the position for broad feature A in the case of the scraped sample.}	
	\end{figure}

	Figure \ref{fig:XPS} shows the valence band photoemission spectra collected using Al $K_\alpha$ and He \textsc{ii} radiations. The Al $K_\alpha$ spectra exhibit three distinct spectral features, A, B and C appearing at about 1.8 eV, 4 eV and 7 eV binding energy, respectively, along with zero intensity at the Fermi level. Similarly, He \textsc{ii} spectra also exhibit all three features at similar energy positions with different relative intensities. Interestingly, the intensity in both spectra appears only above 1 eV binding energy which is smaller than the experimentally obtained band gap\cite{Enhessari2012, Agui2011}. Valence band of MnTiO$_3$ is primarily composed of Mn 3$d$ and O 2$p$ hybridized states, with negligibly small contribution from Ti 3$d$ states. A marked increase in the relative intensity of features C and B compared to feature A is observed while going from Al $K_{\alpha}$ to He \textsc{ii} spectra. This enhancement is primarily due to larger photoionization cross-section ratio of O 2$p$ states to Mn 3$d$ states at lower photon energies \cite{YEH19851}. Therefore, features A can be primarily attributed to states having Mn 3$d$ character, while features B and C have O 2$p$ character.

	\begin{figure}[t]
	\centering
	\includegraphics[width=0.48\textwidth]{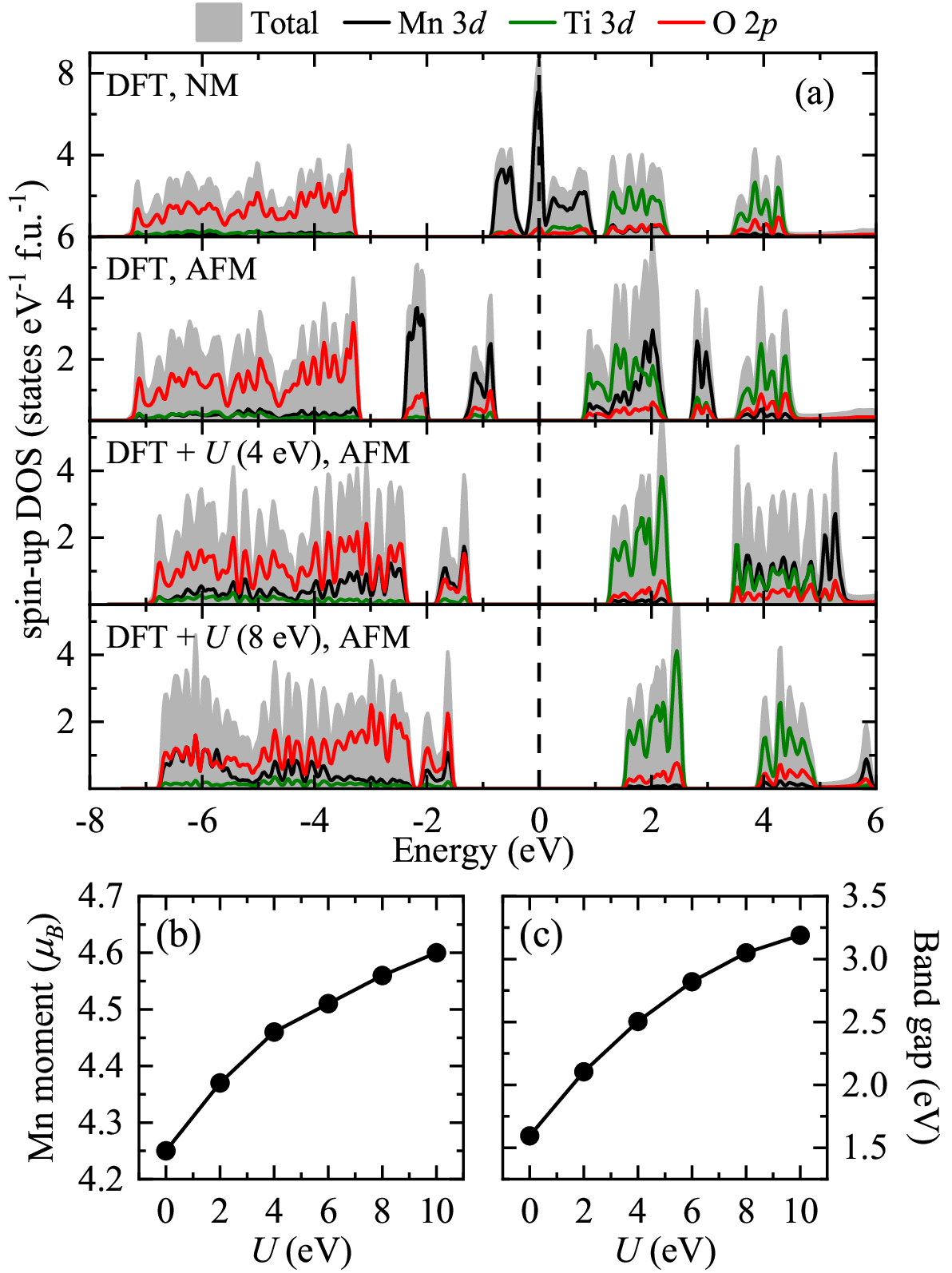}
	\caption{\label{fig:DFT2}  (a) Total and partial DOS for MnTiO$_3$ from DFT and DFT+$U$ calculations. Fermi level (zero energy) is set at middle of the gap. Variation of (b) magnetic moment and (c) band gap with $U$ in the AFM phase.}
	\end{figure}

	Since photoemission spectroscopy is a highly surface sensitive technique, thus we compare the Al $K_{\alpha}$ spectra from the fractured and subsequently scraped sample surface in the inset of figure \ref{fig:XPS}. Scraping leads to spectral shift of about 0.6 eV towards higher binding energy and feature A appears as a hump like feature as shown by an arrow. Increased broadening in spectral features may appear due to larger scattering at the scraped surface. Similar energy shift and broadening observed in core level spectra \cite{Maurya_2017} confirm the rigid band shift in the case of the scraped sample. Due to strong charging effect at lower temperatures, reliable data could not be collected for the AFM phase.   

	For further insight into the electronic structure, we show the results of DFT and DFT+$U$ calculations in Fig. \ref{fig:DFT2}. The nonmagnetic (NM) phase exhibits a metallic state in clear contrast to the insulating behaviour. Forced spin-degeneracy within NM DFT calculation gives zero local moment, while the PM phase is described by randomly oriented (disordered) local moments leading to total zero magnetization, thus the NM results can not be compared with the PM phase. The total energy reduces by 2.00 eV/f.u. in the ferromagnetic (FM) phase and by 2.57 eV/f.u. in the AFM phase with respect to the NM phase, suggesting AFM to be the ground state. While going from the NM to AFM phase (second panel), the energy position of the O 2$p$ and Ti 3$d$ bands remain similar, whereas Mn 3$d$ states are redistributed in four different energy regions. In the occupied region, crystal field split Mn $t_{2g}$ and $e_{g}$ band centers appear at about -2.2 eV and -1 eV energy, respectively, while their exchange split counterparts appear at about 1.5 eV and 3 eV energy. Crystal field splitting of about 1 eV and exchange-splitting of about 4 eV are clearly evident (see Fig. S1 of the SM). The large exchange splitting of Mn 3$d$ band results in an insulating gap of 1.6 eV and a magnetic moment of 4.25 $\mu_{B}/$Mn, which are considerably underestimated from experimentally obtained values. These deviations from the experimental values are presumably due to an inadequate description of electron correlation within semi-local DFT methods \cite{DFTcorrection}. 
	
	With the motivation of improving band gap and magnetic moment, we incorporate the effect of on-site electron correlation ($U$) applied to Mn 3$d$ orbitals within the DFT+$U$ \cite{AnisimovLDAU}. The results are shown in two bottom panels of Fig. \ref{fig:DFT2}(a) for varying $U$. The inclusion of small $U$ improves the magnetic moment and band gap while the overall electronic structure does not change much with respect to the total width and energy positions of occupied O 2$p$ states and unoccupied Ti 3$d$ states. It is clearly evident that unoccupied Mn 3$d$ states are pushed at higher energy leading to a dominant contribution of Ti 3$d$ states in the lowest unoccupied band consistent with $x$-ray absorption spectroscopy (XAS) \cite{Agui2011, SWChen_XAS_APL2014}.  No moment was found at Ti site in these calculations suggesting that it does not contribute to the magnetic properties of MnTiO$_3$. The Mn magnetic moment and band gap increase monotonously with increasing $U$, as shown in Fig. \ref{fig:DFT2}(b) and (c), respectively. The experimental values of the magnetic moment and band gap are well captured within these DFT+$U$ calculations with $U$ about 8 eV, confirming the strongly correlated nature. The density of states (DOS) in the occupied region exhibits interesting evolution where Mn 3$d$ states show spectral weight transfer from the highest occupied band appearing at about -2 eV to lower energy at about -6 eV with increasing $U$. This leads to a change in the dominant contribution from Mn 3$d$ to O 2$p$ character for the highest occupied band, which is in sharp contrast to the experimentally observed valence band (Fig. \ref{fig:XPS}) where feature A has primarily Mn 3$d$ character.
	
	Although the results for the AFM phase can not be one-to-one compared with the photoemission spectra and other experimental results in the PM phase, a reasonably good agreement between DFT or DFT+$U$ results in the AFM phase and experimental results of the PM phase have been found \cite{DFTcorrection, PhysRevB.49.14211,PhysRevB.44.943, FTran_PRB2006_LDAU_monooxides, Mandal2019, SYKim_PhysRevLett2018_MnPS3, PhysRevB.103.134407}. Up to now, the electronic structure and effect of electron correlation are investigated within DFT and DFT+$U$ framework in the AFM ground state due to the inaccessibility of the electronic structure in the PM phase. We now turn our focus on the electronic structure of MnTiO$_{3}$ within the DFT+DMFT framework, which provides a realistic description of the PM phase since it accounts for the fluctuation of local moments and has been very successful in describing strongly correlated magnetic systems \cite{PhysRevB.100.245109, MinjaeKim_PRB2015_DMFTSrRuO3, PhysRevLett.87.067205, Haule_PRL2019_MnPS3, Jonsson_PRB022_MAXphaseDMFT, Tianye_PRB2022_CrI3}. 

 	\begin{figure}[t]
		\centering
		\includegraphics[width=0.48\textwidth]{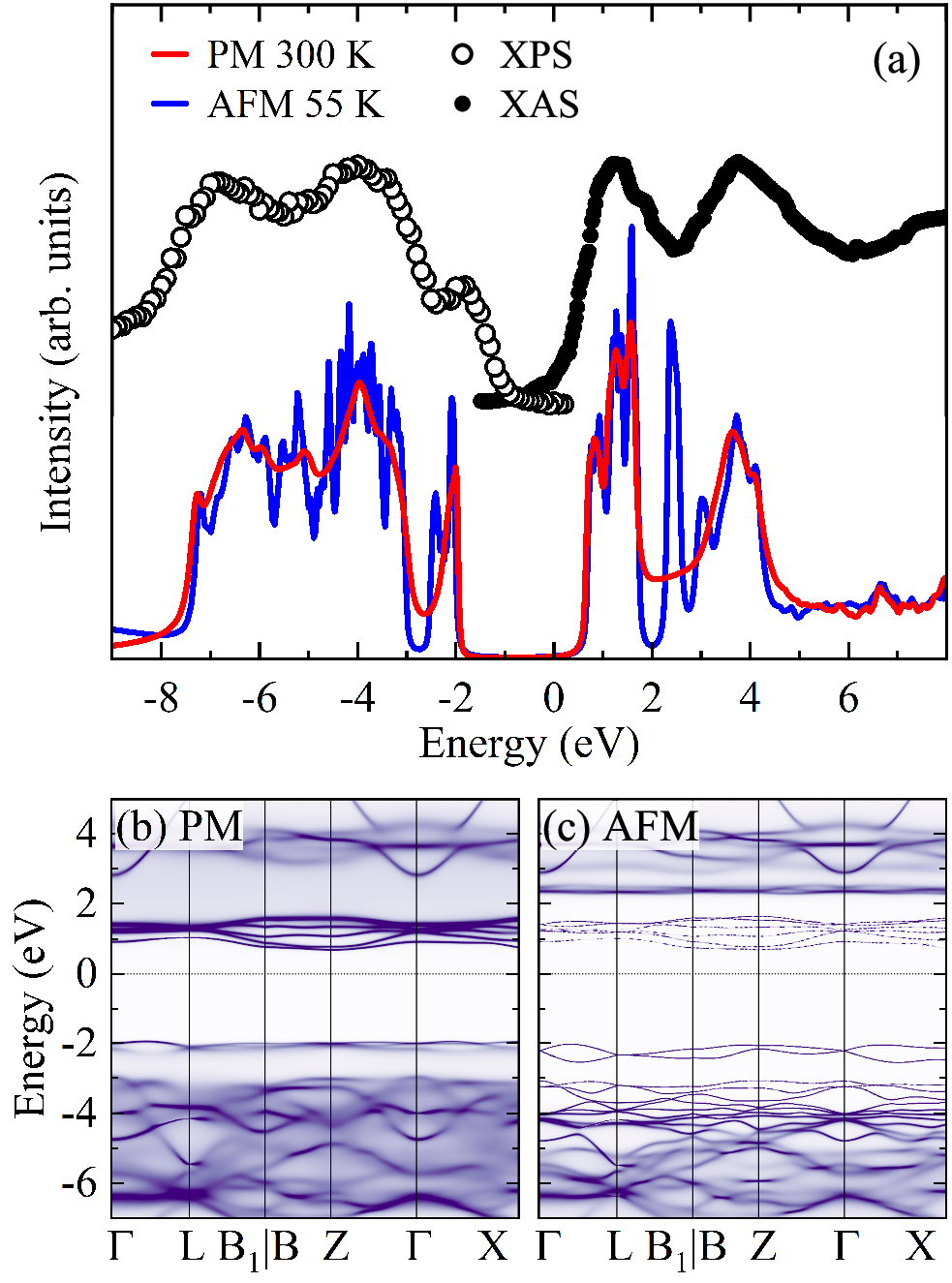}
		\caption{\label{fig:DMFT1}  (a) DFT+DMFT spectral function for MnTiO$_{3}$ in PM phase (red line) and AFM state (blue line) with $U$ = 6.0 eV and ${J_H}$ = 0.9 eV. Room temperature Al $K_\alpha$ spectra (XPS) (open circles) and XAS spectra (closed circles, reproduced from Ref. \cite{Agui2011}). Momentum resolved spectral function in (b) PM and (c) AFM phase.}
	\end{figure} 
    Figure \ref{fig:DMFT1} shows the DFT+DMFT spectral function calculated for the PM phase with $\beta$ $\sim$ 38.68 eV$^{-1}$ ($T$ $\sim$ 300 K), Hubbard-$U$ = 6.0 eV and Hund's $J_{H}$ = 0.9 eV (see Fig. S2 and S3 in SM for results with other $U$ and $J_{H}$ values).  It is to note here that the NM DFT+$U$ calculations do not produce a gap even for $U$ as large as 10 eV. Interestingly, PM DFT+DMFT produces an insulating gap of about 2.6 eV, in close agreement with the band gap obtained from $x$-ray spectroscopic experiments \cite{Agui2011}. The spectral function primarily exhibits three features in the occupied part consistent with the valence band photoemission spectra and two broad features in the unoccupied part consistent with the XAS spectra. For comparison, we show the Al $K_{\alpha}$ valence band spectra (Fig. \ref{fig:XPS}) and O $K$-edge XAS spectra (reproduced from Ref. \cite{Agui2011}) in the same figure. For clarity, experimental spectra have been vertically shifted. For a direct comparison, the calculated spectral function has been shifted to match feature A of the valence band spectra, thereby aligning the Fermi levels. The O $K$-edge XAS spectra (first absorption peak appearing at $\sim$ 531 eV photon energy \cite{Agui2011}) is plotted on the binding energy scale after subtracting O 1$s$ binding energy (529.6 $\pm$ 0.2 eV). Since the photoionization and photoabsorption cross-sections are not taken into account in the calculations, we only compare the energy width and position of spectral features with experimental spectra. Overall, an excellent agreement with the experimentally obtained occupied valence band (Al $K_{\alpha}$ spectra ) as well as the unoccupied conduction band (XAS spectra) is obtained. Atomic contributions to various features in the spectral function (Fig. S4 of the SM) confirm that the -2 eV feature has predominantly Mn 3$d$ character,  consistent with valence band results discussed earlier. The nominal occupancy of the Mn 3$d$ band obtained from CTQMC impurity solver is about 5.09 and the local magnetic moment has been calculated using $\sum_{i} 2P_i | S^{z}_{i} |$ where $P_i$ and $| S^{z}_{i} |$ represent the probability and absolute magnetic moment, respectively, for the $i^{th}$ state out of the 1024 possible atomic states in the impurity solver. The calculated local magnetic moment of 4.75 $\mu_B$/Mn is close to about 5 $\mu_B$/Mn obtained from the experimental paramagnetic effective moment of 5.92 $\mu_B$/Mn  \cite{Stickler, YAMAUCHI1983_spinflop}. 
	
	To understand the electronic structure of low temperature AFM phase, we also show the AFM DFT+DMFT spectral function for T $\sim$ 55 K ($\beta$ $\sim$ 211 eV$^{-1}$) in Fig. \ref{fig:DMFT1}(a). The spectral function does not change significantly across magnetic phase transition but exhibits sharper features in the AFM phase. The band gap in the AFM phase is similar to that obtained in the PM phase, as clearly seen in the figure. The nominal occupancy is 4.91 and 0.17 for the up and down spin, respectively, leading to the static magnetic moment of 4.74 $\mu_B$/Mn, which is almost saturated to the PM local moment. The -2 eV feature in the AFM DFT+DMFT spectral function exhibits two peak like structure as observed in the AFM DFT+$U$ result. It is to note here that the dominant atomic character remains Mn 3$d$ for the highest occupied valence band (Fig. S5 of SM) which was substantially small in DFT+$U$ results for large $U$. An additional sharp feature with Mn 3$d$ character (Fig. S5 of SM) at about 2.4 eV is observed in the conduction band, while a broad hump like feature is observed in PM phase.
	In Fig. \ref{fig:DMFT1}(b) and (c), we show the momentum resolved spectral function for PM phase and AFM phase, respectively, calculated along the specific $k$-path (Fig. \ref{fig:fig1}(a)). The dispersion of spectral function in these two phases are strikingly similar. The dispersion is more diffusive/incoherent in the case of PM phase due to large fluctuating moment than in the case of magnetically ordered AFM phase (more like a mean-field) which can be well described within the band picture \cite{PhysRevB.100.245109}. Clearly, the 2.4 eV feature in the AFM phase has become most diffusive in the PM phase, giving rise to a hump like structure in spectral function, as seen in Fig. \ref{fig:DMFT1}(a). The sharp features at $\sim$ 2.4 eV and $\sim$ 3 eV correspond to the unoccupied Mn $t_{2g}$ and $e_{g}$ states, respectively (Fig. S5 of SM). As the ordered moment in AFM phase reduces by a few percent ($\sim$ 3\%) with increasing temperature, these features become incoherent/diffused (Fig.~S6 of SM) due to increased thermal fluctuations disrupting the ordering of magnetic moments. Such sharpening of the spectral features with magnetic ordering have also been observed in other studies \cite{PhysRevB.100.245109, Haule_PRL2019_MnPS3}. The overall comparison discussed here suggests that the long-range magnetic ordering in MnTiO$_3$ has a negligibly small influence in the occupied part. XAS or inverse photoemission spectroscopy at low temperatures can further validate the change observed here in the unoccupied part across the magnetic phase transition.
	
%\section*{CONCLUSION}
	To summarize, we have investigated the electronic structure of ilmenite MnTiO$_{3}$ using core level and valence band photoemission spectroscopy along with theoretical calculations within DFT, DFT+$U$ and DFT+DMFT frameworks. Spectral lineshape of Mn 2$p$ and Ti 2$p$ core levels are very similar to those observed in MnO and TiO$_{2}$, respectively, suggesting similar environment/octahedral connectivity and strength of various interaction parameters. Multiplet structures were clearly observed in the Mn 2$p_{3/2}$ spectra presumably due to high-quality and clean sample surface. Room temperature valence band spectra corresponding to the PM phase suggests a wide gap insulating state which can not be described within NM DFT. Although AFM DFT+$U$ correctly produces the band gap and magnetic moment, the contribution to the highest occupied band changes from Mn 3$d$ to O 2$p$ for large $U$. The experimental spectra are accurately reproduced by DFT+DMFT calculation for the PM phase. The obtained band gap and PM moment are consistent with $x$-ray spectroscopic experiments and magnetic susceptibility measurements, respectively. The long-range AFM ordering has minimal influence on the spectral function leading to a similar band gap while the ordered moment is almost saturated at 4.74 $\mu_B$/Mn in close consistency with the neutron diffraction measurements. Overall, our study provides detailed insight into the electronic properties of MnTiO$_{3}$ and highlights the importance of electron correlation and fluctuating magnetic moment to accurately describe its electronic structure in the paramagnetic phase.

\acknowledgments
    We thankfully acknowledge R. Bindu, IIT Mandi for helpful discussion. We acknowledge the support of Central Instrumentation Facility and HPC Facility at IISER Bhopal. Support from DST-FIST (Project No. SR/FST/PSI-195/2014(C)) is also thankfully acknowledged.

\end{document}

\title{Supplemental Material for ``Influence of anti-ferromagnetic ordering and electron correlation on the electronic structure of MnTiO$_3$"
	}% Force line breaks with \\
	
\author{Asif Ali}
%\email{asifa@iiserb.ac.in}
\affiliation{Department of Physics, Indian Institute of Science Education and Research Bhopal, Bhopal Bypass Road, Bhauri, Bhopal 462066 India}%

\author{R. K. Maurya}
%\email{asifa@iiserb.ac.in}
\affiliation{Department of Physics, Indian Institute of Science Education and Research Bhopal, Bhopal Bypass Road, Bhauri, Bhopal 462066 India}%

\author{Sakshi Bansal}
%\email{asifa@iiserb.ac.in}
\affiliation{Department of Physics, Indian Institute of Science Education and Research Bhopal, Bhopal Bypass Road, Bhauri, Bhopal 462066 India}%

\author{B. H. Reddy}
\altaffiliation[Presently at ]{Department of Physics, Government College (A), Rajahmundry 533105, India}
%\email{harinath@gcrjy.ac.in}
\affiliation{Department of Physics, Indian Institute of Science Education and Research Bhopal, Bhopal Bypass Road, Bhauri, Bhopal 462066 India}%

\author{Ravi Shankar Singh}
\email{rssingh@iiserb.ac.in}
\affiliation{Department of Physics, Indian Institute of Science Education and Research Bhopal, Bhopal Bypass Road, Bhauri, Bhopal 462066 India}%
		
\maketitle	

\subsection{Electronic structure calculation details }
%\subsection{Theoretical calculation details}
 DFT calculations were performed using the full-potential linearized augmented plane-wave method as implemented in \textsc{wien2k} \cite{WIEN2k, *Wien2k2019}. Generalized gradient approximation of Perdew-Burke-Ernzerhof exchange-correlation functional \cite{PBEPRL1997} and experimental lattice parameters were used for all the calculations. 1000 $k$-points were considered within the first Brillouin zone while energy and charge convergence criteria were set to 0.1 meV and 10$^{-4}$ electronic charge per formula unit (f.u.), respectively. The G-type AFM ordering was considered for calculations of the AFM phase \cite{Shirane1959_neutron}. The DFT+$U$ method \cite{AnisimovLDAU} was adopted to account for on-site electron correlation at Mn site. Fully charge self-consistent DFT+DMFT calculations were performed using eDMFT code \cite{HauleDMFTPRB2010}. 
All five Mn 3\textit{d} orbitals were considered as correlated orbitals. The impurity problem was solved by continuous-time quantum Monte Carlo (CTQMC) solver \cite{Haule_CTQMC_PRB2007} with 10$^8$ Monte Carlo steps in each DMFT iteration. The density-density form of the Coulomb interaction was employed and double counting method was set to \textit{exactd}. The maximum entropy method was used for the analytic continuation of the self-energy from the imaginary frequency axis to the real frequency axis. \cite{JARRELL1996_maxent}. In all the calculations, 5000 $k$-points were used to calculate the density of states (DOS).

\subsection{Mn 3$d$ spin-projected DOS from DFT AFM calculation }
\begin{figure}[H]
	\centering
	\includegraphics[width=.65\textwidth]{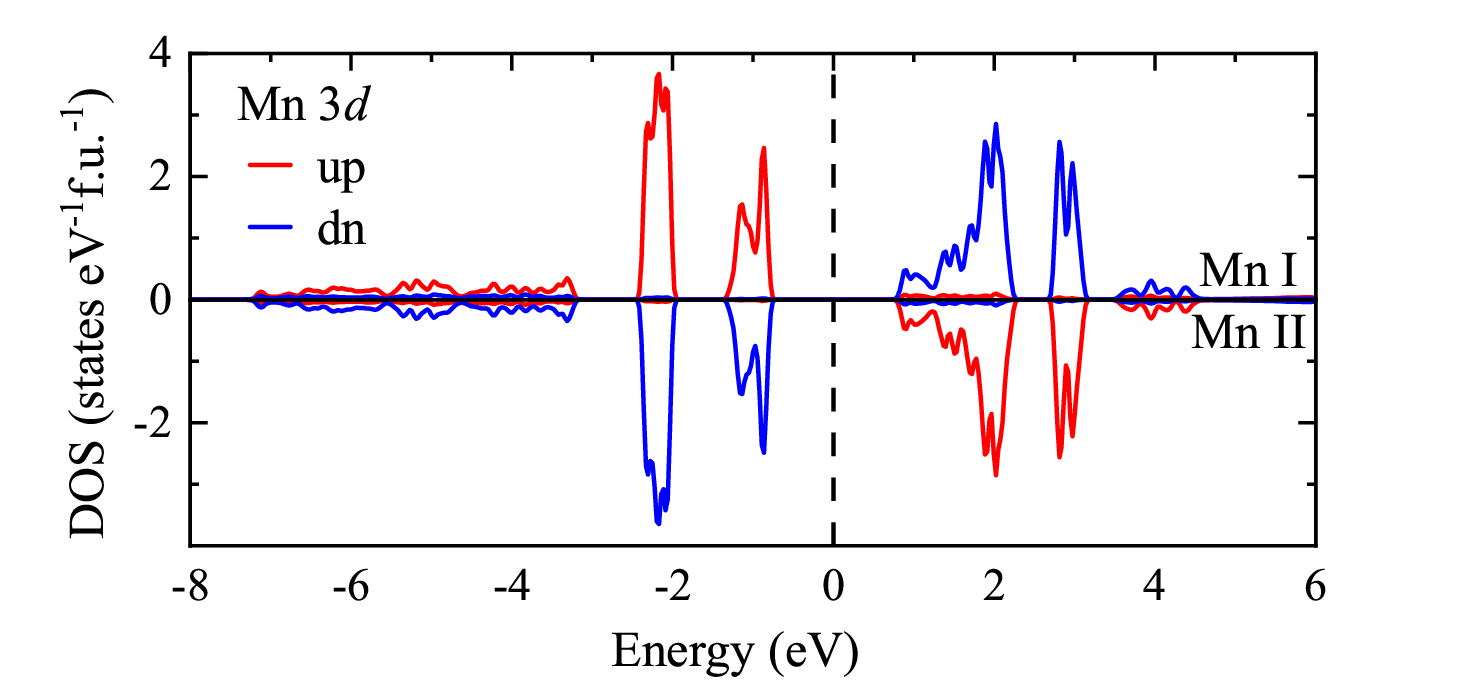}
	\caption{Up (red) and down (blue) Mn 3$d$ partial DOS corresponding to two neighbouring Mn sites. Mn-I on the positive $y$-axis and Mn-II on the negative $y$-axis.} \label{fig:FigS1}
\end{figure}

Fig. \ref{fig:FigS1} shows the spin-resolved Mn 3$d$ PDOS for the anti-ferromagnetically aligned Mn sites (Mn-I and Mn-II) calculated from DFT calculation for AFM phase. Both exhibit well separated up and down spin channels, leading to completely filled up-spin channel and completely empty down-spin channel for Mn-I and vice versa for Mn-II. This confirms the high-spin configuration for Mn$^{2+}$ ion, where all five electrons have parallel spins to maximize the total spin. Crystal field splitting of about 1 eV and exchange-splitting of about 4 eV are clearly evident.

\subsection{DFT + DMFT spectral function within varying $U$ and $J_{H}$ }
\begin{figure}[H]
	\centering
	\includegraphics[width=.85\textwidth]{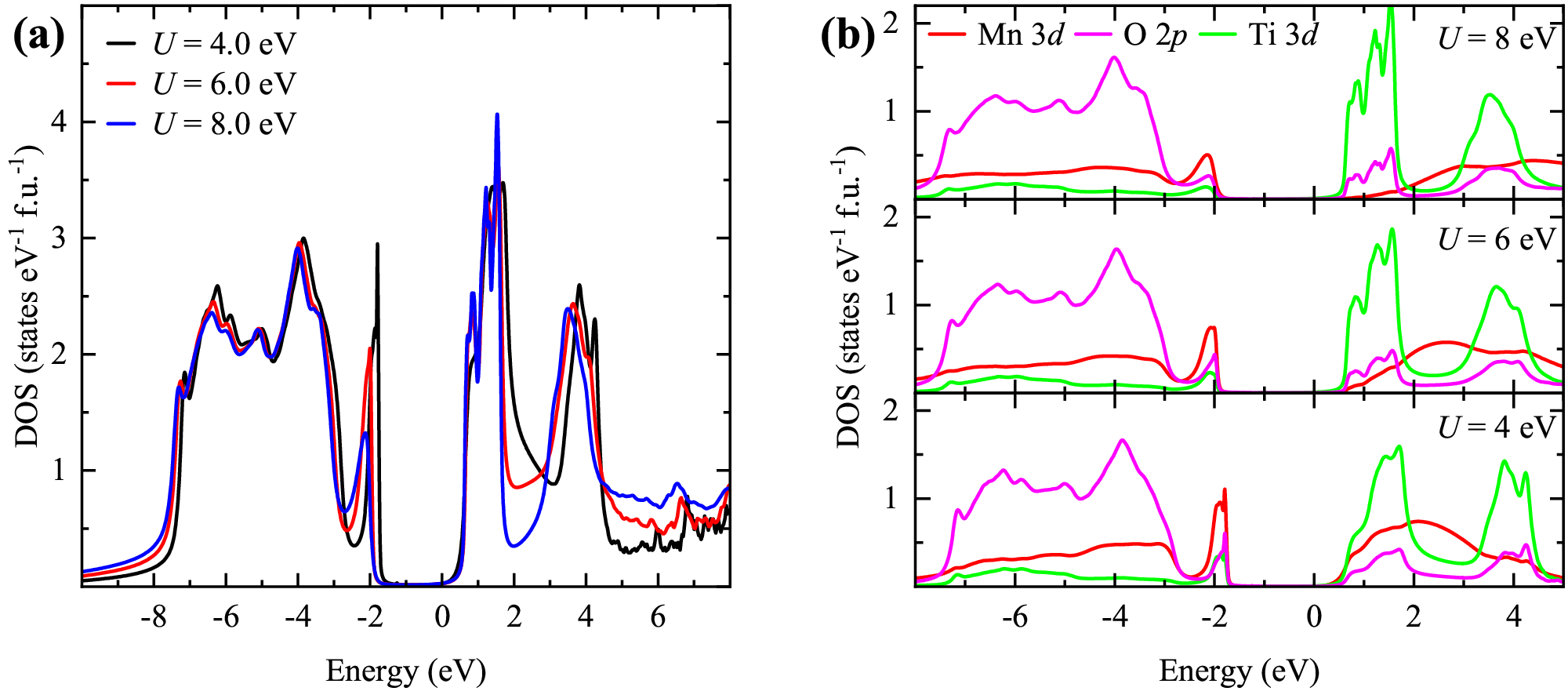}
	\caption{(a) DFT+DMFT spectral functions with varying $U$ and fixed $J_H$ = 0.9 eV for MnTiO$_{3}$ in the paramagnetic phase. (b) Atom resolved spectral functions for different $U$.} \label{fig:FigS2}
\end{figure}

Fig. \ref{fig:FigS2}(a) shows the DFT+DMFT spectral functions with varying $U$ and fixed $J_H$ = 0.9 eV for MnTiO$_{3}$ in the paramagnetic phase. All the features remain at very similar positions while the intensity of the -2 eV feature slightly reduces with increasing $U$. Atom resolved spectral functions for $U$ = 4 eV, 6 eV and 8 eV are shown in the top, middle and bottom panels of Fig. \ref{fig:FigS2}(b), respectively. Although the band gap (Mn $d$ - Ti $d$ gap) remains similar, the Mn $d$ - Mn $d$ gap increases with increasing $U$.

\begin{figure}[H]
	\centering
	\includegraphics[width=.85\textwidth]{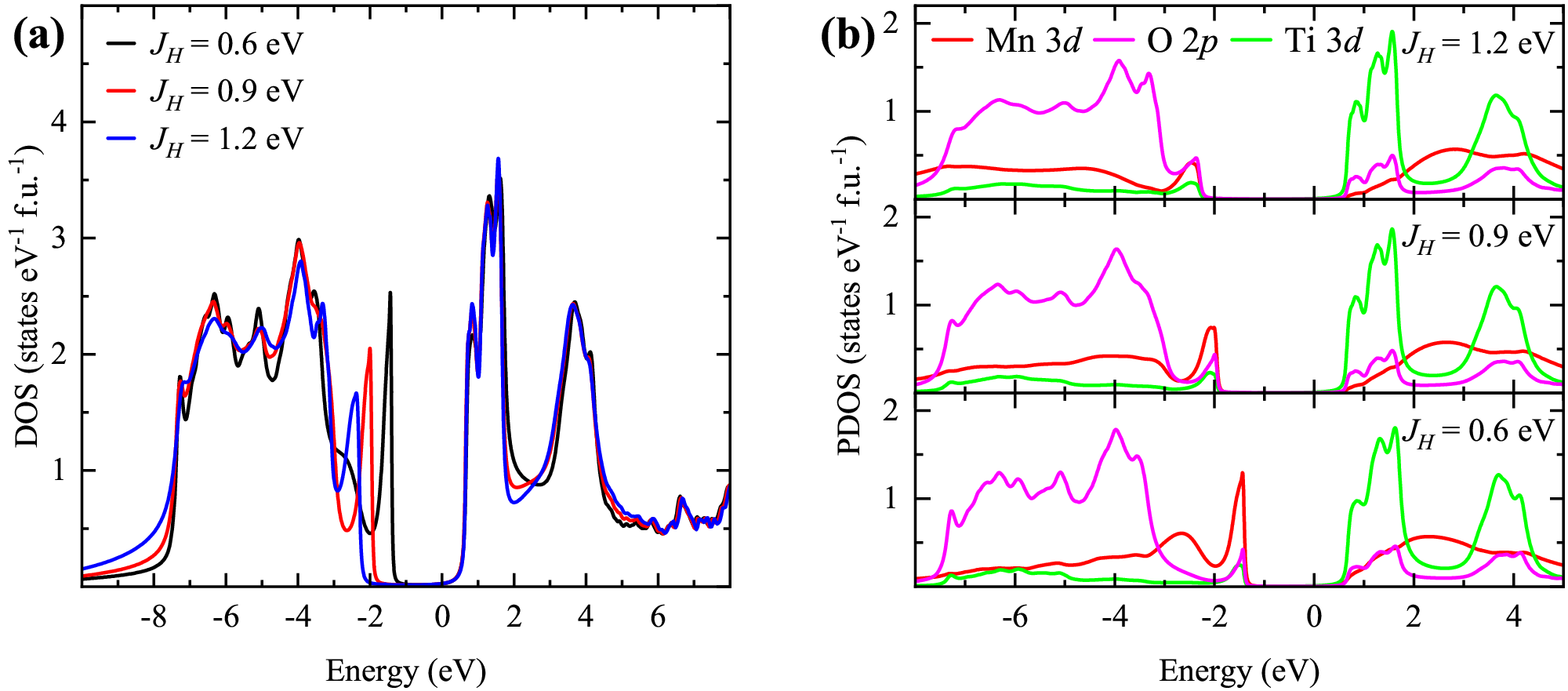}
	\caption{(a) DFT+DMFT spectral functions with fixed $U$ = 6 eV and varying $J_{H}$ for MnTiO$_{3}$ in the paramagnetic phase. (b) Atom resolved spectral functions for $J_{H}$ = 1.2 eV, 0.9 eV and 0.6 eV .} \label{fig:FigS3}
\end{figure}

Fig. \ref{fig:FigS3}(a) shows the DFT+DMFT spectral functions with fixed $U$ = 6 eV and varying $J_{H}$ for in the paramagnetic phase. All the features remain at very similar positions except for the -2 eV feature, which moves away from the Fermi level thereby increasing the band gap with increasing $J_H$. Atom resolved spectral functions for $J_{H}$ = 1.2 eV, 0.9 eV and 0.6 eV are shown in the top, middle and bottom panels of Fig. \ref{fig:FigS3}(b), respectively.  Increasing $J_{H}$ reduces the contribution of Mn 3$d$ states in feature A. For large $J_{H}$ = 1.2 eV, the dominant character changes from Mn 3$d$ to O 2$p$ as opposed to the experimental observation. 

\subsection{Projected spectral function from DFT+DMFT}
\begin{figure}[H]
	\centering
	\includegraphics[width=.85\textwidth]{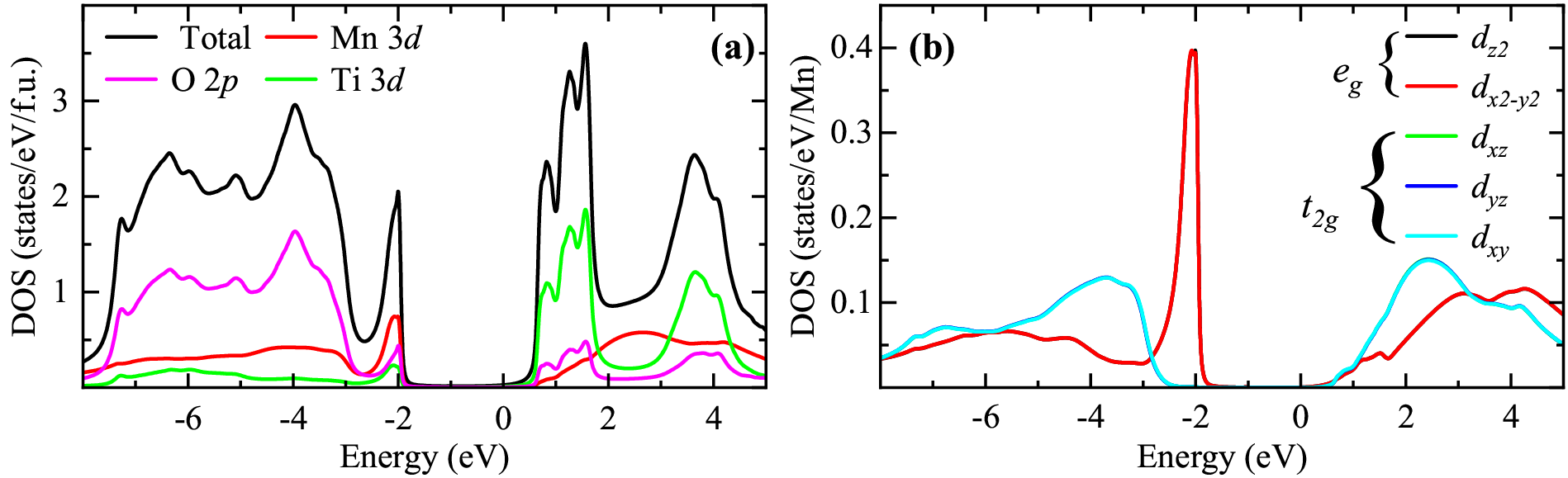}
	\caption{(a) Total and atom resolved spectral functions obtained from DFT+DMFT calculation for $U$ = 6 eV and $J_H$ = 0.9 eV for MnTiO$_3$ in the paramagnetic phase.  (b) Mn 3$d$ orbital projected spectral functions.}\label{fig:FigS4}
\end{figure}

Total and atom resolved spectral functions obtained from DFT+DMFT calculation for $U$ = 6 eV and $J_H$ = 0.9 eV for MnTiO$_3$ in the paramagnetic phase is shown in Fig. \ref{fig:FigS4}. Dominant Mn 3$d$ character is seen in the -2 eV feature consistent with the valence band photoemission spectra. Mn 3$d$ exhibits broad structure in energy below -3 eV (occupied part) as well as above 1 eV (unoccupied part) consistent with highly diffusive bands in dispersion shown in Fig 4 (b) of the main text. Mn 3$d$ orbital projected spectral functions exhibit $t_{2g} - e_{g}$ like crystal-field splitting of the Mn 3$d$ band where each orbital is almost half-filled. The nominal occupancy obtained from the CTQMC is 5.09.

\begin{figure}[h]
	\centering
	\includegraphics[width=.85\textwidth]{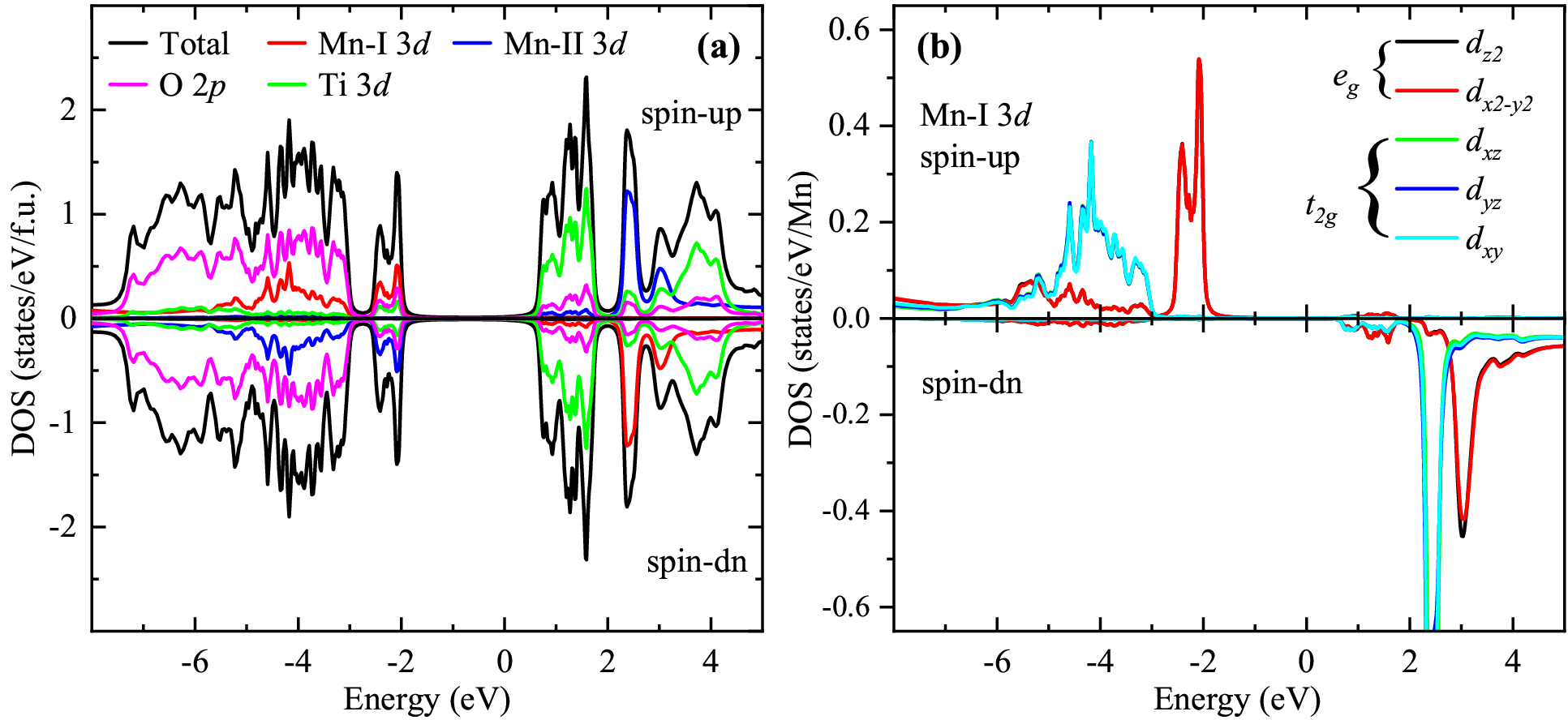}
	\caption{(a) Total and atom resolved spectral  functions obtained from DFT+DMFT calculation for $U$ = 6 eV and $J_H$ = 0.9 eV for MnTiO$_3$ in the anti-ferromagnetic phase. (b) The orbital projected spectral functions for MnI-3$d$.} \label{fig:FigS5}
\end{figure}

 Fig. \ref{fig:FigS5}(a) shows the total and atom resolved spectral functions obtained from DFT+DMFT calculation for $U$ = 6 eV and $J_H$ = 0.9 eV for MnTiO$_3$ in the anti-ferromagnetic phase. The feature at about 2.4 eV in the unoccupied region has dominant contribution from the Mn 3$d$ states. The orbital projected spectral functions for MnI 3$d$ (Fig. \ref{fig:FigS5}(b)), suggest that a small spin-down contribution appears in the occupied region. The nominal occupancy obtained from CTQMC is 4.91 and 0.17 for the up and down spin, respectively, consistent with the order magnetic moment of 4.74 $\mu_{B}$/Mn.

\subsection{Influence of temperature and magnetic ordering on spectral function}
\begin{figure}[h]
	\centering
	\includegraphics[width=.65\textwidth]{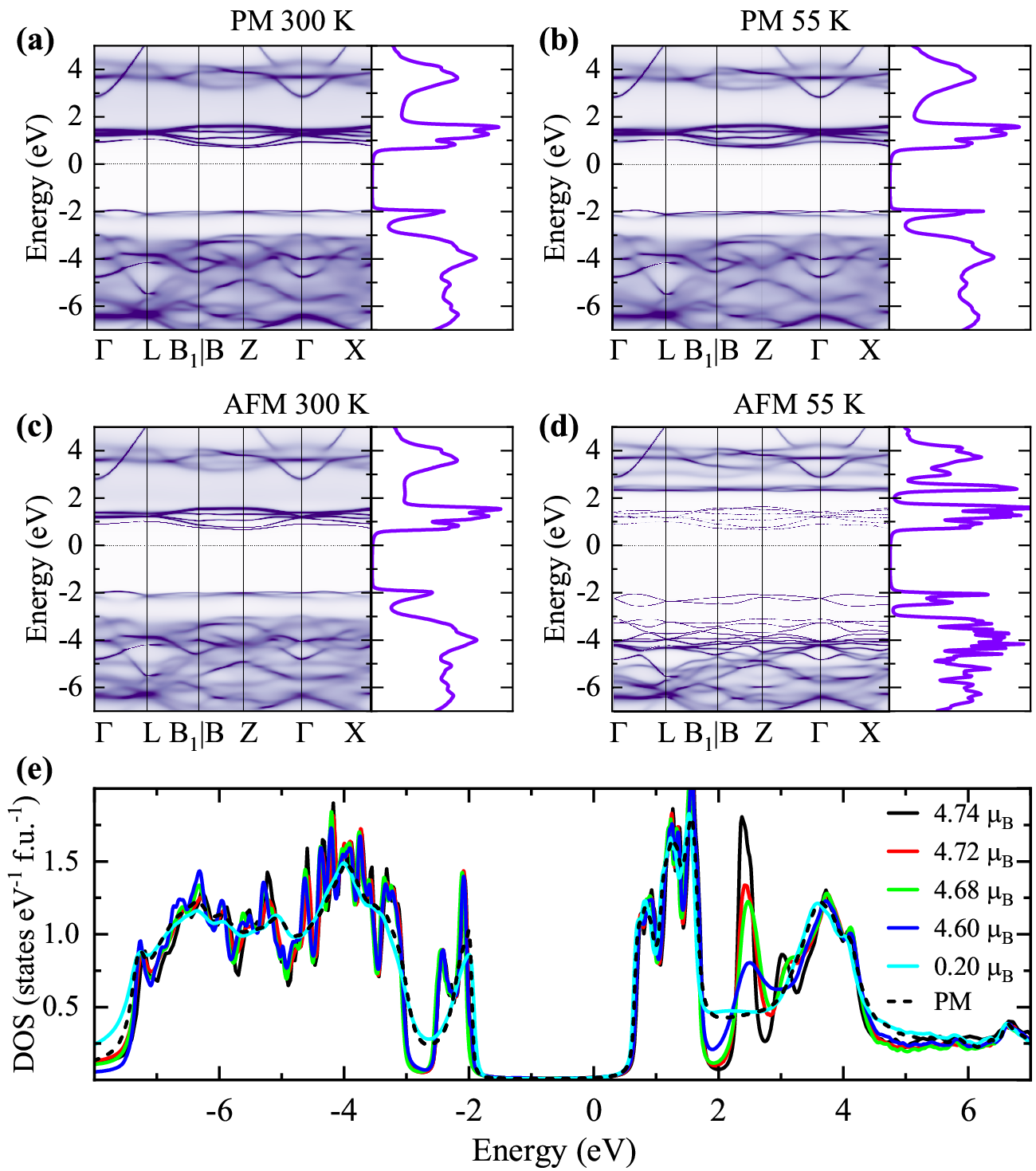}
	\caption{DFT+DMFT calculated $k$-resolved and $k$-integrated spectral functions in the AFM and PM phases at different temperatures. PM phase at (a) 300 K and (b) 55 K and AFM phase at (c) 300 K and (d) 55 K. (e) $k$-integrated spectral functions of the AFM phase at different temperature. Dash line shows spectral function in the PM phase at 300 K.} \label{fig:FigS6}
\end{figure}

DFT+DMFT calculated $k$-resolved and $k$-integrated spectral functions in the AFM and PM phases at different temperatures have been compared in Fig. \ref{fig:FigS6}. Fig. \ref{fig:FigS6} (a)  and (d) are same as shown in the main text. Spectral function in the PM phase at 300 K (Fig. \ref{fig:FigS6} (a)) remain very similar down to 55 K (Fig. \ref{fig:FigS6} (b)). As expected, spectral function in the AFM phase with negligibly small ordered moment (0.2 $\mu_{B}$/Mn) at 300 K is similar to that of the PM phase (both 300 K and 55 K). Spectral function in the AFM phase at 55 K with ordered moment of 4.74 $\mu_{B}$/Mn exhibits sharp coherent bands (more like a mean-field) which can be well described within the band picture. A careful look suggests that the Mn $d$ bands ($t_{2g}$ and $e_g$ both) in the occupied (-5 eV to -2 eV) as well as unoccupied (at 2.4 eV and 3.0 eV) become coherent while other states remain very similar. In Fig. \ref{fig:FigS6} (e), we further show the evolution of $k$-integrated spectral function obtained within AFM DFT+DMFT with increasing temperatures (which leads to decreasing ordered moment). Spectral function for 55 K, 100 K, 150 K and 175 K with ordered moment of 4.74, 4.72, 4.68 and 4.60 $\mu_{B}$/Mn has been shown by black, red, green and blue lines respectively. The sharpness of spectral features at $\sim$ 2.4 eV and 3.0 eV decreases as the moment decreases and these bands (unoccupied $t_{2g}$ and $e_g$ respectively) become incoherent/diffusive when the ordered moment decreases by a few percent. 300 K AFM (cyan lines) and 300 K PM (black dashed lines) results have also been shown for comparison suggesting the disappearance of 2.4 eV and 3 eV features due to the diffusive nature of these bands at higher temperature.
 
\newpage
%\bibliography{biblography}
